\documentstyle[emulateapj]{article}

\lefthead{Bekki Kenji}
\righthead{Group-cluster merging}

\begin{document}
\title{Group-cluster merging and the formation of starburst galaxies}

\author{Kenji Bekki} 
\affil{Astronomical Institute, Tohoku University, Sendai, 980-8578, Japan \\
email address: bekki@astroa.astr.tohoku.ac.jp}

\begin{abstract}

 A significant fraction of clusters of galaxies are observed to have substructure,
 which implies  that merging between clusters and subclusters is
 a rather common physical process of cluster formation.
 It still remains unclear how cluster merging affects 
 the evolution of  cluster member galaxies. 
 We report the results of numerical simulations, which show the dynamical evolution of a 
 gas-rich late-type spiral in a merger between a small group of galaxies and
 a cluster. The simulations demonstrate that time-dependent tidal gravitational field
 of the  merging excites non-axisymmetric structure of the galaxy,
 subsequently drives efficient transfer of gas to the central region,
 and finally triggers a secondary starburst. 
 This result provides not only a new mechanism of starbursts but also
 a close physical relationship between the emergence of starburst galaxies and the
 formation  of substructure in clusters.
We accordingly  interpret  post-starburst galaxies located near
 substructure of the Coma cluster as one observational example  indicating  the global tidal effects
 of group-cluster merging. 
 Our numerical results  furthermore suggest
 a causal link between the observed excess of blue  galaxies in 
 distant clusters and
 cluster virialization  process through hierarchical merging of subclusters.

\end{abstract}


\section{Introduction}
  Recent observational studies on morphology, structure, and kinematics
  of nearby and distant clusters of galaxies have revealed that 
  a significant fraction of clusters have substructures (e.g., Forman \& Jones 1990;
   Briel et al. 1991; White,  Briel, \& Henry 1993; Escalera et al. 1994; Slezak, Durret, \& Gerbal 1994;
  Colless \& Dunn 1996; L$\rm \acute{e}monon$ et al. 1997).
  Theoretical studies  based largely on numerical simulations 
  suggest that these substructures in clusters 
  result from  cluster formation process  
  through merging of other smaller subclusters  and groups of galaxies
  (Evrard 1990;  Burns et al. 1994; Roettiger, Stone, \& Mushotzky 1998).
  This cluster merging process has been found to affect greatly star formation
  histories of cluster member galaxies (e.g., Caldwell et al. 1993; Caldwell \& Rose 1997).
  Caldwell et al. (1993) and  Caldwell \& Rose (1997) found that
  post-starburst galaxies are located preferentially
  near a secondary peak in the x-ray emission of the Coma cluster, and accordingly
  suggested that merging between the main Coma and a group of galaxies (including NGC 4839)
  plays  a vital role in triggering secondary starbursts in  galaxies.
  Caldwell \& Rose (1997)  furthermore revealed that about 15 \% of early-type galaxies
  in nearby five rich clusters with substructures have clear signs of ongoing or recent
  starbursts, and speculated a causal relation between the formation of starburst galaxies
  and cluster merging process.

  Although  merging between clusters and  subclusters (or groups)
  is observationally suggested to affect greatly  star formation histories
  of cluster member galaxies,
  there have been  no extensive theoretical studies investigating   
  how cluster merging determines  or changes the star formation histories
%  of cluster member galaxies
  in the course of cluster formation.
  In this Letter,  we numerically investigate stellar and gas dynamics of a bulgeless late-type spiral 
  in a merger between two  clusters of galaxies. In particular, we describe how 
  the time-dependent tidal gravitational field of the  merging determines 
  the star formation history of the gas-rich spiral
  and thereby  illustrate  a new mechanism for triggering
  starbursts in cluster environments.
  We here emphasize that not the simple mean tidal gravitational
  field of a cluster but the rapidly changing  
  tidal gravitational field of  group-cluster merging is a necessary ingredient 
  for triggering a starburst in group-cluster merging.
  Accordingly, the present mechanism of starburst galaxy formation is different
  from the previously proposed models which resort to the simple tidal gravitational
  field of clusters (e.g., Byrd \& Valtonen 1990; Byrd \& Henriksen 1996; Moore et al. 1996;
  Moore, Lake, \& Katz 1998).
  Cluster merging can play decisive roles in various aspects of cluster galaxy evolution,
  such as morphological transformation, dynamical heating, tidal truncation of gas replenishment,
  and excitation of nuclear activities in galaxies. 
  We however focus mainly on  star formation histories of  cluster member  galaxies in this Letter.

\section{Model}
  The structure of two clusters in a merger is modeled by using
  the universal density profile  predicted 
  by the standard cold dark matter  cosmogony (Navarro, Frenk,  \& White 1996).
  For convenience, we refer to the smaller (larger) cluster as a group (cluster). 
  Total mass of a  cluster (group) represented by $M_{\rm cl}$
  ($M_{\rm gr}$) is set to be  $2.0 \times 10^{14} \rm M_{\odot}$
  ($5.0 \times 10^{13}  \rm M_{\odot}$). 
  The scale radius and virial one  
  are  127 kpc (80.1 kpc) and  1.16 Mpc (728 kpc), respectively, for the cluster (group).
  The group merges with the cluster 
  with the impact parameter of 254 kpc, the initial relative  velocity of $V_{\rm rel}$, 
  and the initial separation of  1.73 Mpc. 
  We present the results of two models with $V_{\rm rel}=430$ km/s
  and 602 km/s
  (corresponding to 0.5 and 0.7 times the circular 
  velocity at the virial radius of the cluster, respectively).
  Although the strength of the induced starburst depends on $M_{\rm cl}$ and $M_{\rm gr}/M_{\rm cl}$
  in such a way  that starbursts are more likely to be stronger for the model with
  larger $M_{\rm cl}$ and $M_{\rm gr}/M_{\rm cl} \sim 1.0$
  (for $10^{13} \rm M_{\odot} \leq \it M_{\rm gr},M_{\rm cl} 
  \rm \leq 5.0 \times 10^{14}  M_{\odot}$), 
  we here present only the 
  results of the models
  with $ M_{\rm cl} = 2.0 \times 10^{14} \rm M_{\odot}$ and $M_{\rm gr}/M_{\rm cl}=0.25$.
  More details on $M_{\rm cl}$, $M_{\rm gr}$, and $V_{\rm rel}$ dependences are given
  in Bekki (1998).

  We construct the  model of a  bulgeless gas-rich disk 
  by using the Fall-Efstathiou  model (Fall \& Efstathiou 1980).
  Initial mass-ratio of dark matter halo
  to disk stars to disk gas is 20:4:1 for the  disk.
The disk mass  and size are 
6.0 $\times$ $10^{10}$ $ \rm M_{\odot}$ and 17.5 kpc, respectively.
The  exponential disk scale length  
and the maximum rotational velocity for the disk is 
3.5 kpc
and  220  km/s,
respectively.
In addition to the rotational velocity made by the gravitational
field of the disk,
the initial radial and azimuthal velocity
dispersion are given to disk components according
to the epicyclic theory with Toomre's stability  parameter $Q$ (Binney \& Tremaine 1987)
equal to 1.5.
      Collisional and dissipative nature of gas is represented by discrete gas clouds rather
      than by a continuum of gas governed by an equation of state.
      This scheme we adopt is called as `Sticky particle method' (Schwarz 1981).
      In this scheme, gas clouds collide with each other inelastically,
      reduce their relative velocity, and lose their kinetic energy dissipatively.
      The size and the mass for each gas cloud
      are set to be $1.3 \times {10}^{2} $ pc and $1.2 \times {10}^{6} \rm M_{\odot}$,
      respectively. The radial and tangential restitution coefficient for cloud-cloud
      collisions are  set to be 1.0 and  0.0, respectively.
Star formation in  gas clouds is modeled by using
the Schmidt law (Schmidt 1959) with the exponent of  2.0 and 
the threshold gas density (Kennicutt 1989)  of  
0.22 $\rm M_{\odot}$/${\rm pc}^{3}$.
The coefficient of the Schmidt law is chosen such that
the mean star formation rate for the first 1 Gyr evolution
of an isolated disk is $\sim 1$ $\rm M_{\odot}$/yr.
The disk  orbits the center of the group and the initial distance between
the mass center of the group and the disk is 80.1 kpc.
The dependence of star formation history of a disk galaxy on the initial distance
is described in detail by Bekki (1998).

  In order to elucidate more clearly  the importance of time-dependent
  tidal gravitational effects 
  of  group-cluster merging in star formation histories of  galaxies,
  we  assume firstly that only the group has a disk galaxy,
  and secondly that neither  the group nor  the cluster have hot intracluster medium.
  Accordingly a group-cluster merger is composed of a galaxy,  background dark matter of
  a group, and that of a cluster.
  This assumption means that we 
  here neglect completely tidal  effects of galaxy interaction and merging 
  (Barnes \& Hernquist 1992; Moore et al. 1996),
  ram-pressure stripping (Farouki \& Shapiro 1980), and hydrodynamical interaction between galaxies
  and intracluster medium (Evrard 1991),
  and thereby extract $only$ the essential ingredient of the tidal effects of  group-cluster
  merging.
  It should be emphasized here that the present study does not consider physical mechanisms
  of starbursts other than group-cluster merging to be less important: The relative importance
  of each of possibly promising mechanisms for triggering starbursts in clusters
  should be clarified by
  future more  realistic theoretical studies. 
  Thus, although the present numerical study is rather idealized and less realistic in some points,
  it can clearly demonstrate that 
  the tidal gravitational effects of group-cluster merging drives secondary starbursts of galaxies.

 The total particle number is 20000 
 for background dark matter of a cluster, 10000 for that of a group,
 10000 for dark matter halo of a disk, 
 10000 for stellar components of the disk, and 10000 for gaseous
 ones of the disk, in a group-cluster merging.
 The  softening length is 32.2 kpc 
 for  the gravitational interaction between the dark
 matter  components of the cluster  and  those of the group
 (corresponding
 to the mean particle separation at the half mass radius of the cluster),
 0.81 kpc 
 for that between disk stars, gas, and dark matter halo surrounding
 the disk
(the mean particle separation at the half mass radius of the disk),
 and 16.5 kpc for that  
 between dark halo components of the cluster and the group and disk components
 (the mean value of the above two softening lengths).
 We use these three different gravitational softening lengths in order to
 investigate how the global tidal field of group-cluster merging
 affects the local dynamical evolution of the disk in an admittedly self-consistent manner.
 All the simulations have been carried out on the GRAPE board (Sugimoto et al. 1990).

\section{Result}

Figure 1 describes the typical behavior of the formation of a starburst galaxy
in group-cluster merging.
After the group passes through the cluster for the first time,
it suffers from tidal distension and gradually  loses its central concentration 
(the time $T=3.4$ Gyr).
Then, the group becomes widely dispersed owing to strong
global tidal effects of the cluster whereas the cluster does not change so drastically
its initial mass distribution ($T=4.5,7.9$).
Finally, the merger becomes dynamically relaxed to form a new and more massive cluster. 
Substructure of the merger can be clearly seen between $T=3.4$ and $T=7.9$.
These formation processes of substructure in a cluster merger are consistent
reasonably well with those of previous studies (Burns et al. 1994; Roettiger et al. 1998).

Time-dependent tidal gravitational field 
of the merger gives strong non-axisymmetric perturbation to a disk galaxy
in the group 
and subsequently induces a central  stellar bar and outer asymmetric gaseous spiral arms
in the galaxy ($T=3.4$).
These non-axisymmetric structures enhance cloud-cloud collisions
and gaseous dissipation, drive the efficient  transfer
of gas to the central region of the galaxy,
and consequently induce a secondary starburst ($T=4.0$ in Fig.1 and Fig.2).
As is shown in Figure 2,
the strength of the starburst in the present group-cluster merger model is rather moderate compared
with that triggered by major galaxy merging (Mihos \& Hernquist 1996).
The secondary starburst populations form
a central small flat bulge-like component and consequently 
increase the degree of  central mass concentration 
in  the disk ($T=7.9$).
In total, about 72\% of initial gas is consumed by star formation. 
These results clearly demonstrate a physical connection between the emergence of starburst (or post-starburst)
disk galaxies and cluster merging process.

As the galaxy sinks toward
the center of the cluster owing to dynamical friction
in the late phase of the merging, 
the stellar disk suffers from strong dynamical heating
resulting from  tidal gravitational field of the cluster core.
Consequently, the disk
is dynamically thickened and the spiral structure becomes rather inconspicuous.
Furthermore, about 45 \% of dark halo components  of the disk 
are tidally stripped during the merging.
Galactic haloes are generally considered to be gas reservoirs indispensable for supplying
fresh gas and maintaining star formation activity in  disks (Larson, Tinsley, \& Caldwell 1980).
Tidal removal of halo components accordingly could drive the truncation of gas infall and replenishment
and consequently curtail the future star  formation of disk galaxies in group-cluster mergers.
These results imply that global gravitational effects of cluster mergers can
not only drive  morphological transformation of late-type disks but also control
star formation histories of those galaxies.

Thus, the present  numerical results 
imply that group-cluster merging can transform actively star-forming gas-rich
late-type disks into
gas-poor early-type ones with considerably inactive star formation,
principally because global tidal gravitational field of the merging
can greatly affect both galactic star formation histories and morphology.
These accordingly  suggest that 
as a cluster undergoes  hierarchical merging of small groups 
and subsequently becomes more dynamically relaxed and more massive,
the fraction of late-type spirals in the cluster becomes progressively smaller: 
Group-cluster merging provides a causal relation between rapid morphological transformation
of late-type spirals and cluster growth process through merging of subclusters.
%The origin of the so-called density-morphology relation is closely associated
%with hierarchical growth of clusters.
Recent observational studies (Edge \& Stewart 1991;  Smail et al. 1997)
have found a marginal evidence that there is an anticorrelation
between the inferred total mass of a cluster and the fraction of  spirals within the cluster.

\section{Discussion and conclusion}

Remarkable features of the present group-cluster merger model of starburst galaxies 
are the following two.
Firstly, the model demonstrates a physical connection between the presence of starburst (or post-starburst)
galaxies in clusters and the existence of substructure in those clusters.
In particular, in the merger with $V_{\rm rel}=602$ km/s,
the starburst (or post-starburst) galaxy  remains  well outside the cluster and 
located near the developed substructure for a few Gyr (See Figure 3).
This result strongly suggests that the origin of  widespread post-starburst galaxies located near substructure
of the Coma cluster (Caldwell et al. 1993)
is due principally to tidal gravitational effects of a group-cluster merger
with rather large relative velocity.
The origin of the observed larger fraction of post-starburst galaxies in nearby clusters with obvious double 
structure (Caldwell \& Rose 1997) can be also explained by the tidal effects of cluster mergers.

Secondly, galaxies with starbursts (or post-starbursts) neither have conspicuous  tidal tails 
nor are severely distorted in the present model.
This provides  an important implication on 
the origin of  the Butcher-Oemler effect (Butcher \& Oemler 1978),
 which means that higher redshift clusters of galaxies 
 have a larger fraction of blue galaxies than lower redshift counterparts.
 Recent morphological studies 
 by the Hubble Space Telescope and grand-based telescopes
 (Couch et al. 1994; Lavery \& Henry 1994;  Couch et al. 1998)
  have revealed that 
  about the half of distant blue galaxies  appear to be interacting or merging,
  which strengthens the importance of galaxy interaction and merging in the star 
  formation histories of disk galaxies.
  However, the remaining half of the galaxies, though discernibly distorted and having
  asymmetric feature, neither have remarkable tidal tails  nor 
  are severely disturbed (Couch et al. 1998).
  The starbursts triggered by  group-cluster merging can give a natural explanation
  for these distant blue galaxies.
  We here stress  that not all of distant blue galaxies are observationally
  revealed to experience the past starbursts (Abraham et al. 1996; Barger et al. 1996):
  Infall of field galaxies with active star formation can equally explain the
  origin of blue galaxies with largely unperturbed morphology in distant clusters.

 We conclude that time-dependent  tidal gravitational field of group-cluster merging,
 the importance of which has been completely neglected in previous studies,
 induces secondary starbursts.
 Galaxy interaction and merging (e.g., Barnes \& Hernquist 1992), 
 galaxy  harassment (Moore et al. 1996), 
 and  environmental effects of clusters (Farouki \& Shapiro 1980; Byrd \& Valtonen 1990;  Evrard 1991)
 are all suggested to play decisive roles in the formation of starburst galaxies.
 Then,  which physical process dominates the formation of
 starburst galaxies in cluster environments?
The present study predicts that only one time group-cluster merging can trigger starbursts 
in a significant fraction of group member galaxies nearly simultaneously, 
principally because the tidal effect of the merging
is global and thus  can induce the formation of non-axisymmetric structure  in  the galaxies.
Accordingly,  
one of observational tests to assess the relative importance of group-cluster merging
is to investigate whether or not
starburst galaxies are developed 
nearly simultaneously in an ongoing cluster merger.

\acknowledgments
We are grateful to the referee James A. Rose for valuable comments,
which greatly contribute to improve the present paper.
K.B thanks to the Japan Society for Promotion of Science (JSPS)
Research Fellowships for Young Scientist.

\clearpage
%%%%%%%%%%%%%%%%%%%%%%% Figure Captions

\begin{figure}
\caption[]{
 Morphological evolution of a group, a cluster, and a disk galaxy
 in a group-cluster merger for
 the model with $M_{\rm cl} = 2.0 \times 10^{14} \rm M_{\odot}$,
 $M_{\rm gr} = 5.0 \times 10^{13} \rm M_{\odot}$,
 and $V_{\rm rel} = 430$  km/s.
 Each frame measures 4.73 Mpc (35 kpc) for the upper (lower) panels,
 and the time, indicated in the upper left corner of each panel, 
 is in units of Gyr.
 The group is assumed to enter the cluster from the right side in this figure. 
 In the upper four panels, background dark matter components are shown in cyan 
 for the cluster and in magenta for the group.
 A cross in each panel represents the position of the disk in the group for each time.
 In the lower four panels, 
 stellar components (stars and  new stars that are originally gaseous components
 and converted into stellar ones by star formation)
 and gaseous ones are shown
 in magenta and cyan, respectively.
 We here do not intend to display dark halo components of the disk in order to  show more clearly 
 the morphological evolution of disk components.
 \label{fig-1} }
 \end{figure}

\begin{figure}
\caption[]{
 The time evolution
 of global star formation rate of a disk galaxy 
 in  group-cluster merging  for the models with $V_{\rm rel}=430$ km/s (magenta).
 Here star formation rate relative to that of an isolated disk
 is shown for each time.
 Accordingly this figure  describes to what degree the star formation of the disk
 in group-cluster merging is enhanced in comparison with that of isolated disk evolution.
 The result of a disk galaxy within an isolated group 
 with $M_{\rm gr}$  = $5.0 \times 10^{13} \rm M_{\odot}$  
 and that of a disk galaxy within an isolated cluster 
 with $M_{\rm cl}$  = $2.0 \times 10^{14} \rm M_{\odot}$ are also
 shown in green and in cyan, respectively,
 in order that tidal effects of time-dependent gravitational field
 of group-cluster merging can be more clearly  
 demonstrated to play a vital role in triggering a starburst.
 For these isolated group and cluster models, a spiral 
 orbits the center of the isolated 
 cluster (group) with the pericenter distance of 127 (80.1) kpc. 
 As is shown in this figure, the rapidly changing tidal gravitational
 field in group-cluster merging can more greatly destabilize the spiral
 and trigger a stronger starburst than isolated group and cluster models.
 \label{fig-2} }
 \end{figure}

\begin{figure}
\caption[]{
 Morphology  of a  group-cluster merger at the time $T=3.4$ Gyr  
 in the model with 
 $V_{\rm rel} = 602$ km/s.
 Dark matter components of the cluster and those  of the group 
 are shown in cyan  and in magenta, respectively.
 The frame of this figure measures 5.67 Mpc,
 and a  cross represents the position of a disk galaxy initially within the  group. 
 Starburst galaxies (or post-starburst ones)
 are more likely to remain  well outside the cluster and located 
 within the diffusely distributed dark matter components
 of the destroyed group for $3.4 \leq T \leq 7.9$ Gyr in this group-cluster merging
 with rather large initial relative velocity.
 \label{fig-3} }
 \end{figure}


\newpage





\begin{thebibliography}{}
%\begin{thebibliography}



\bibitem[Abraham et al.  1996]{ab96}
Abraham, R. G., Smecker-Hane, T. A., Hutchings, J. B., Carlberg, R. G.,
Yee, H. K. C., Ellingson, E., Morris, S., Oke, J. B., \& Rigler, M.
1996, \apj,   471, 694


\bibitem[Barger et al. 1996]{ba96}
Barger, A. J., Arag$\rm \acute{o}$n-Salamanca, A., Ellis, R. S.,
Couch, W. J., Smail, I., \& Sharples, P. M. 
1996, \mnras,  279, 1

\bibitem[Barnes \& Hernquist 1992]{bh92}
Barnes, J., \& Hernquist, L. 1992, \araa, 30, 705 


\bibitem[Bekki  1998]{be98}
Bekki, K. 1998, in preparation


\bibitem[Binney \& Tremaine 1987]{bt87}
Binney, J., \& Tremaine, S. 1987,  Galactic Dynamics, Princeton; Princeton
Univ. Press.


\bibitem[Briel et al. 1991]{br91}
Briel, U. G., Henry, J. P., Schwarz, R. A., B$\ddot{\rm o}$hringer, H.,
Ebeling, H., Edge, A. C., Hartner, G. D., Schindler, S., Tr$\ddot{\rm u}$mper, J.,
\& Voges, W.  1991, A\&A,  246, L10

\bibitem[Burns et al. 1994]{bu94}
Burns, J. O., Roettiger, K., Ledlow, M., \& Klypin, A.
1994, \apj,  427, L87

\bibitem[ Butcher \& Oemler 1992]{bu92}
Butcher, H. \& Oemler, A. 
1978, \apj, 219, 18


\bibitem[Byrd \& Valtonen 1990]{by90}
Byrd, G. \& Valtonen, M. 
1990, \apj,  350, 89

\bibitem[Byrd \& Henriksen  1996]{by96}
Byrd, G. \& Henriksen, M. J. 
1996, \apj,  459, 82

\bibitem[Caldwell et al 1993]{ca93}
Caldwell, N., Rose, J., Sharples, R. M., Ellis, R. S., \& Bower, R. G.
1993, \aj, 106, 473

\bibitem[Caldwell \& Rose 1997]{ca97}
Caldwell, N., \& Rose, J.
1997, \aj, 113, 492

\bibitem[Colless \& Dunn 1996]{co96} 
Colless, M., \& Dunn, A. M.,
1996, \apj, 458, 435

\bibitem[Couch et al. 1994]{co94}
Couch, W. J., Ellis, R. S., Sharples, R. M., \& Smail, I.
1994, \apj, 430, 121

\bibitem[Couch et al. 1998]{co98}
Couch, W., Barger, A. J., Smail, I., Ellis, R. S., \& Sharples, R. M.
1998, \apj,   497, 188


%\bibitem[Dressler et al. 1997]{dr97}
%Dressler, A., Oemler, A.,  Couch, W. J., Smail, I., Ellis, R. S.,
%Barger, A., Butcher, H., Poggianti. B. M., \&  Sharples, R. M.
%1997, \apj,  490, 577
%\bibitem[Ellis et al. 1997]{ell97}
%Ellis, R. S., Smail, I., Dressler, A., Couch, W. J., Oemler, A.,
%Butcher, H., \& Sharples R.M. 1997, \apj, 483, 582

\bibitem[Edge \& Stewart 1991]{ed91}
Edge, A. C., \& Stewart, G. C.
1991, \mnras, 252, 428

\bibitem[Escalera et al.  1994]{es94}
Escalera, E., Biviano, A., Girardi, M., Giuricin, G., Mardirossian, F.,
Mazure, A., \& Mezzetti, M.
1994, \apj,   423, 539 

\bibitem[Evrard 1990]{ev90}
Evrard, A. E. 1990, \apj, 363, 349


\bibitem[Evrard  1991]{ev91}
Evrard, A. E. 
1991, \mnras,   248, 8-10p 


\bibitem[Fall \& Efstathiou 1980]{fe80}
Fall, S. M., \& Efstathiou, G. 1980, \mnras, 193, 189


\bibitem[Farouki \& Shapiro 1980]{fa80}
Farouki, R. \& Shapiro, S. L. 
1980, \apj,   241, 928

\bibitem[Forman \& Jones 1990]{fj90}
 Forman, W., \& Jones, C. 1990,
 in Clusters of Galaxies (eds. Oergerle, W. R.,  Fichett, M. J.,
 \&  Danly, L.), p257 (Cambridge University Press, 1990)


%\bibitem[Icke  1985]{ic85}
%Icke, V.
%1985, A\&A,  144, 115

\bibitem[Kennicutt 1989]{ken89}
Kennicutt, R. C. 1989, \apj, 344, 685



\bibitem[Larson et al. 1980]{lar80}
Larson, R. B., Tinsley, B. M., \& Caldwell, C. N. 1980,
\apj, 237, 692

\bibitem[Lavery \& Henry 1994]{la94}
Lavery, R. J. \& Henry, J. P.
1994, \apj,   426, 524

\bibitem[L$\rm \acute{e}monon$ et al. 1997]{le97}
L$\rm \acute{e}$monon, M., Pierre, M., Hunstead, R., Reid, A., Mellier, Y.,
\& B\"ohringer, H.
1997, A\&A,   326, 34 

%\bibitem[Lubin et al. 1998]{lu98}
%Lubin, M. L., Postman, M., Oke, J. B., Ratnatunga, K. U., Gunn, J. E., Hoessel, J. G.,
%Schneider, D. P.
%1998, \apj,  in press, astro-ph/9804286 
%\bibitem[Mihos   1995]{mi95}
%Mihos, J. C. 1995, \apj, 438, L75

\bibitem[Mihos \& Hernquist 1996]{mi96}
Mihos, J. C., \& Hernquist, L.
1996, \apj, 464, 641

\bibitem[Moore et al. 1996]{mo96}
Moore, B., Katz, N., Lake, G., Dressler, A., \& Oemler, A.
1996, \nat, 379, 613

\bibitem[Moore et al. 1998]{mo98}
Moore, B., Katz, N., \& Lake, G.
1998, \apj, 495, 139 

\bibitem[Navarro et al. 1996]{na96}
Navarro, J. F., Frenk, C. S., \& White, S. D. M.
1996, \apj, 462, 563

\bibitem[Roettiger et al. 1998]{ro98}
Roettiger, K., Stone, J. M., \& Mushotzky, R. F. 
1998, \apj,  493, 62

\bibitem[Schmidt 1959]{sch}
Schmidt, M. 1959, \apj, 344, 685

%\bibitem[Sandage 1961]{sa61}
%Sandage, A. 1961, The Hubble Atlas of Galaxies

\bibitem[Schwarz 1981]{sch81}
Schwarz, M. P. 1981, \apj, 247, 77


\bibitem[Smail et al. 1997]{sm97}
Smail, I., Ellis, R. S., Dressler, A., Couch, W., Oemler, A., Sharples, R. M.,
\& Butcher, H.
1997, \apj,  479, 70


\bibitem[Sugimoto et al. 1990]{sug90}
Sugimoto, D., Chikada, Y., Makino, J., Ito, T., Ebisuzaki, T., \& 
Umemura, M. 1990, \nat, 345, 33

\bibitem[White et al. 1993]{wh93}
White, S. D. M.,  Briel, U. G., \& Henry, P.
1993, \mnras, 261, L8

%\bibitem[van Dokkum  et al. 1998]{va98}
%van Dokkum, P. G., Franx, M., Kelson, D. D., Illingworth, G. D.,
%Fisher, D. \& Fabricant, D.
%1998, \apj,  in press, astro-ph/9801190
%\bibitem[Wielen 1977]{wie77}
%Wielen, R. 1977, \aap, 60, 263
%\bibitem[Zabludoff et al.  1996]{za96}
%Zabludoff, A. I., Zaritsky, D., Lin, H., Tucker, D., Hashimoto, Y., Shectman, S. A.,
%Oemler A., \& Kirshner, R. P. 
%1996,  \apj,   466, 104









%\bibitem[]{}
%\bibitem[]{}
%\bibitem[]{}
%\bibitem[\protect\citename{}]{}
%\bibitem[\protect\citename{}]{}
%\bibitem[\protect\citename{}]{}

\end{thebibliography}
\end{document}